\documentclass{aa} 
\usepackage{aalongtable,lscape}
\usepackage{natbib}
\usepackage{graphicx,longtable}
\usepackage{txfonts}
\begin{document}
   \title{HD 11397 and HD 14282 - Two new barium stars?
   \thanks{Based on spectroscopic observations collected at the
           European Southern Observatory (ESO), within the
           Observat\'orio Nacional ON/ESO and ON/IAG agreements,
           under FAPESP project n$^{\circ}$ 1998/10138-8.}}

  \subtitle{}

  \author{L. Pomp\'eia
          \inst{1}
     \and
          D.M. Allen \inst{2}
     }
     \offprints{L. Pomp\'eia}

     \institute{IP\&D, Universidade do Vale do Para\'iba, Av. Shishima Hifumi, 2911, S\~ao,
          Jos\'e dos Campos, 12244-000 SP, Brazil\\
	  \email{pompeia@univap.br}
    \and
          Centre for Astrophysics
      Research, STRI and School of Physics, Astronomy and
      Mathematics, University of Hertfordshire, Hatfield, UK. \\
          \email{d.moreira-allen@herts.ac.uk}
	  }

   \date{Received; accepted}

  \abstract
   {We have performed a detailed abundance analysis of the content of $s$-process 
   elements of two dwarf stars with suspected overabundace of those elements. Such stars 
   belong to a special kinematic sample of the solar neighborhood, with peculiar 
   kinematics and different chemical abundances when compared to ``normal" disk stars.}
   {We aim to define if those stars can be identified as barium stars, based on
   their $s$-process elements abundances, and their classification, i.e., if they share 
   their chemical profile with strong or mild barium stars. We also intend to shed 
   light on the possible origins of the different kinds of barium stars.}
   {Spectra have been taken by using the FEROS spectrograph at the 1.52m telescope of ESO,
   La Silla. Abundances have been derived for 18 elements, by matching the synthetic
   profile with the observed spectrum.}
   {We have found that HD 11397 shows a mild enhancement for most of the $s$-process elements
as well as for some $r$-process elements. This star seems to share its abundance
profile with the mild Ba-stars. Although showing some slight chemical anomalies for Y, 
Sr, Mo, and Pb, HD 14282 depicts a chemical pattern similar to the normal stars with
slight $s$-process enhancements.
}
{}


   \maketitle
%

\section{Introduction}
Some fields in astrophysics are particularly complex due to the 
interconnection of many different and, at the same time, important factors. 
Barium stars certainly belong to one of those fields. These are chemically peculiar
stars with enhancements of $s$-process elements and carbon, with 
luminosities below the threshold for the onset of thermal pulses in the AGB 
(Asymptotic Giant Branch) phase. They were identified for the first time by 
Bidelman \& Keenan (1951) as peculiar red giants, and since then many studies 
have been performed in order to better understand their nature and
origin \citep[e.g.][]{mc84,lamb88,luck91,lamb93,nor94,jor98,lia00}.
$s$-process elements are produced along the thermal instability phase of the 
AGB (TP-AGB) \citep[see e.g.][]{ib91} at the interpulse period triggered by the 
{$\rm ^{13}C(\alpha,n)^{16}O$} reaction, and marginally during the convective 
thermal pulses by the {$\rm ^{22}Ne(\alpha,n)^{25}Mg$} reaction 
\citep[e.g.][]{wal97,bus99}. The s-process nucleosynthesis takes place slightly 
above the He-rich, C-rich shell, and during thermal pulses, material enriched 
with $s$-process and carbon ashes is mixed to the H-rich envelope by the third 
dredge-up. As a consequence of this process, the atmospheric abundances of 
TP-AGB stars slowly change with time, with increasing $s$ and C content, 
eventually forming a carbon star.

As Ba stars are in an evolutionary stage before the AGB, their surface abundances 
are puzzling. Two hypotheses have been suggested to explain their chemical 
peculiarity. The first and most accepted is the mass-transfer scenario within 
a binary system by an AGB star to an unevolved (less massive) companion 
\citep{mc80}. Along the AGB, stars develop strong stellar winds, with 
mass-loss rates, ${\dot M}$, from 10$^{-7}$ to 10$^{-5}$ M$_{\odot}$ yr$^{-1}$, 
contaminating the environment with their material \citep{bus01}. Therefore 
binary companions near TP-AGB stars are expected to inherit large amounts 
of this material and change their atmospheric carbon and $s$-process elements 
abundances. In the second hypothesis, Ba stars are formed from ISM clouds 
previously contaminated by AGB activity (e.g. Beveridge \& Sneden 1994). 
Studies about the binary incidence in Ba stars have been extensively performed 
and have strengthened the first hypothesis (McClure 1984, 
McClure \& Woodsworth 1990, Jorissen \& Mayor 1988, North \& Duquennoy 1992,
Karakas et al. 2000). Jorissen et al. (1998) for example studied the radial 
velocity of a sample of Ba stars, with mild or strong peculiarities. They
have found that 35/37 of the strong and 34/40 of the mild Ba stars are 
binaries, ``compatible with the hypothesis that all the observed stars 
are binary systems". Those results strongly support the binary origin 
for the Ba stars (see Pols et al. 2003 and Jorissen \& Van Eck 2005 
for a detailed discussion), although special cases might exist such as 
the chemically peculiar stars from $\omega$ Cen (Norris \& Da Costa 1995, 
Smith et al. 2000).

McClure et al. (1980) and Sneden et al. (1981), based on the work by 
Warner (1965), divided the Ba stars according to their chemical anomalies 
in strong and mild Ba stars. The classification index takes into account 
the Ba II 4554{\rm \AA} line strength, and ranges from 1 to 5. Strong 
Ba stars are more contaminated by $s$-process elements and carbon and 
have an index of $\sim$ 3 to 5; for the mild Ba stars the index is 
$\lesssim$ 2. The process (or processes) by which a normal star becomes a 
strong or a mild Ba star have been discussed by some authors but is still lacking 
a definite conclusion \citep[e.g.][]{pols03,lia03,jor05,smi07}. The orbital 
period of the binary system, the $s$-process efficiency during the AGB phase 
of the more massive star, the type and efficiency of the mass-transfer mechanism,
 the metallicity of the stars, and the mass of the future Ba star 
 which affects the dilution degree of the transferred material, may all 
 play a role in the final abundances. Jorissen et al. (1998) have found 
 that Ba stars with very similar orbital periods show different $s$-enhancements, 
 indicating that other parameters are involved. They suggest that mild, 
 strong and Pop.II CH stars (which show very large $s$-enhancements) 
 correspond to a sequence of increasingly older and more metal-deficient 
 populations. Smiljanic et al. (2007) claim that mild and strong Ba stars 
 show similar iron content, and their difference is probably 
 connected to the neutron exposure during the $s$-process operation. 
 Higher neutron exposures would produce strong Ba stars, while mild Ba 
 stars have accreted matterial exposed to weaker neutron exposures.

In the present work we report chemical abundances for two stars from the 
bulgelike sample of Michel Grenon (Grenon 1999, 2000), HD 11397 and HD 14282. 
Those stars belong to a kinematically selected sample of the solar 
neighborhood, the bulgelike stars (Pomp\'eia et al. 2002, 2003). The 
bulgelike stars have very eccentric orbits, with eccentricities e $>$ 0.25 
and small pericentric distances with R$_{\rm p}$ $\leq$ 3.5 kpc. Therefore 
they were probably born near the galactic bulge (Pomp\'eia et al. 2003, 2008). 
The chemical abundances of the bulgelike stars have been previously studied 
(Pomp\'eia et al. 2003) and their Ba, Zr, La and Y content have been determined. 
HD 11397 and HD 14282 have shown abundance anomalies in their $s$-process content, 
with enhanced [Ba/Fe] and [La/Fe] ratios when compared to the other bulgelike stars 
(see Fig. 8 of Pomp\'eia et al. 2003). HD 14282
also shows overabundant [Zr/Fe] ratios when compared to the other bulgelike stars.
In order to properly classify 
the present stars and to perform a full analysis of their neutron-capture
elements profile, we have inferred their abundances for C, N, La, Ba, Nd, Sm, Sr, 
Zr, Y, Mo, Ru, Ce, Pr, Gd, Dy, Hf and Pb, and for the $r$-process element Eu.

This work is divided as follows: in Sect. 2 we describe the 
observations and reduction procedures, the stellar parameters determination
and the model atmospheres are described in Sect. 3, the abundance 
calculations and their results are given in Sect. 4, in Sect. 5 we discuss our 
results, and in Sect. 6 we give a summary of the paper.


\section{Observations and stellar parameters}
Sample stars have been observed at the 1.52m telescope of ESO 
(European Southern Observatory), La Silla, in September 1999. 
The spectra were obtained using the FEROS spectrograph (Fiber-fed 
Extended Range Optical Spectrograph) with wavelength 
range 356 to 920 nm and a resolution of R = 48,000. The detector 
is a back-illuminated CCD with 2048 X 4096 pixels of 15 $\mu$m size. 
Reductions were performed using the DRS (online data reduction system 
of FEROS) and for a subsequent reduction, the TELLURIC, CONTINUUM, 
RVIDLINES and DOPCOR tasks of the IRAF package were applied.

\begin{table*}
\caption{Photometric and spectroscopic parameters for HD 11397 and HD 14282.}
\label{table:1}
\centering
\begin{tabular}{ccccccc}
\hline\hline
Name & log g $_{Spec}$ & log g$_{Hip}$ & T$_{Spec}$  & [Fe/H]$_{Spec}$ & [Fe/H]$_{Gen}$ & $\xi$(\rm kms$^{-1}$)\\
\hline
HD 11397     &  4.15 $\pm$ 0.2 &  4.34  &  5400 $\pm$ 100  &  -0.72 $\pm$ 0.15  & -0.78  & 0.80 $\pm$ 0.15  \\
HD 14282     &  3.70 $\pm$ 0.2 &  3.91  &  5800 $\pm$ 100  &  -0.60 $\pm$ 0.15  & -0.59  & 1.00 $\pm$ 0.15  \\
\hline
\end{tabular}
\end{table*}

Precise stellar parameters are fundamental for an acurate inference of the 
chemical abundances. Chemical abundances are particularly sensitive to the 
temperature of the stellar atmosphere ({T$_{\rm eff}$) and the surface gravity 
of the star (log g). Two other stellar parameters also play important role 
in the abundance determination: the metallicity of the star ([Fe/H]) and the 
microturbulence velocity ($\xi$). First guesses for the stellar parameters 
were inferred from photometric data and distances from Hipparcos mission. 
The final stellar parameters were calculated as follows: temperatures were derived 
requiring that Fe~I lines with different excitation potentials give the 
same iron abundance; gravities and metallicities were inferred by forcing 
the agreement between Fe~I and Fe~II abundances; microturbulence velocities 
were calculated by requiring no slope in the [Fe/H] vs. EW (equivalent width) 
plot. A detailed description of the entire procedure is given in Pomp\'eia 
et al. (2008, in preparation). The adopted model atmospheres are an updated 
version of the plane-parallel MARCS model atmospheres  with standard 
composition \citep{gus03}. The final parameters and respective uncertainties
are given in Table 1.

\begin{table}[ht!]
\caption{$\log\epsilon$(X) and [X/Fe] for HD 11397 and HD 14282. The solar abundances 
used and their sources are shown in columns 6 and 7. The numbers in parenthesis are the 
errors on the last decimals. The estimated stellar parameters for the two stars are: HD 11397 - 
T$_{eff}$ = 5400K, log g = 4.15 dex, [Fe/H] = -0.72 dex and $\xi$ = 0.80 kms$^{-1}$, HD 14282 - 
T$_{eff}$ = 5800K, log g = 3.70, [Fe/H] = -0.60 dex, $\xi$ = 1.00 kms$^{-1}$.  }
{\scriptsize
\label{medxfe}
   $$ 
\setlength\tabcolsep{3pt}
\begin{tabular}{ccccccrclc}
\hline
\noalign{\smallskip}
 &&  \multicolumn{2}{c}{HD 11397} && \multicolumn{2}{c}{HD 14282} && \multicolumn{2}{c}{Sun} \\
\cline{3-4} \cline{6-7} \cline{9-10}\\
el && $\log\epsilon$(X) & [X/Fe] && $\log\epsilon$(X) & [X/Fe] && $\log\epsilon_\odot$(X) & ref \\
\noalign{\smallskip}
\hline
\noalign{\smallskip}
 C     && 8.14(18) &  0.34(12) &&  8.28(18) &  0.36(12) && 8.52(6) & 1 \\
 N     && 7.50(26) &  0.30(22) &&  7.46(33) &  0.14(30) && 7.92(6) & 1 \\
Sr I   && 2.87     &  0.62     &&  3.07     &  0.70     && 2.97(7) & 1 \\
Sr II  && 2.87(21) &  0.62(17) &&  3.02(21) &  0.65(16) && 2.97(7) & 1 \\
 Y     && 2.17(16) &  0.65(07) &&  2.21(16) &  0.57(7)  && 2.24(3) & 1 \\
Zr I   && 2.36     &  0.48     &&  2.55     &  0.55     && 2.60(2) & 1 \\
Zr II  && 2.59(19) &  0.71(11) &&  2.53(18) &  0.53(10) && 2.60(2) & 1 \\
Mo     && 1.52(26) &  0.32(22) &&  1.72(26) &  0.40(22) && 1.92(5) & 1 \\
Ru     && 1.96(26) &  0.84(22) &&  1.54(20) &  0.30(16) && 1.84(7) & 1 \\
Ba     && 2.36(18) &  0.95(11) &&  2.04(18) &  0.51(11) && 2.13(5) & 1 \\
La     && 1.25(17) &  0.84(08) &&  0.74(18) &  0.21(11) && 1.13(3) & 2 \\
Ce     && 1.71(16) &  0.73(08) &&  1.19(16) &  0.09(08) && 1.70(4) & 3 \\
Pr     && 0.81(25) &  0.87(25) &&  0.37(25) &  0.31(25) && 0.66(15)& 4 \\
Nd     && 1.48(16) &  0.75(07) &&  0.81(17) & -0.04(07) && 1.45(1) & 5 \\
Sm     && 0.83(18) &  0.54(11) &&  0.58(16) &  0.17(09) && 1.01(6) & 6 \\
Eu     && 0.35(22) &  0.55(15) &&  0.33(21) &  0.41(14) && 0.52(1) & 7 \\
Gd     && 0.77(22) &  0.37(16) &&  0.66(19) &  0.14(12) && 1.12(4) & 1 \\
Dy     && 0.99(22) &  0.51(17) &&  0.55(21) & -0.05(15) && 1.20(6) & 8 \\
Hf     && 1.11(22) &  0.95(19) &&  0.58(21) &  0.30(16) && 0.88(8) & 1 \\
Pb     && 1.83(45) &  0.60(44) &&  2.05(41) &  0.70(39) && 1.95(8) & 1 \\
\noalign{\smallskip}
\hline
 \end{tabular}
   $$ 
}
References on solar abundances: 1 - \citet{gs98}; 2 - \citet{law01a}; 
3 - \citet{PQ00}; 4 - \citet{L76}; 5 - \citet{Hartog03}; 6 - \citet{B89}; 
7 - \citet{law01b}; 8 - \citet{BL93}.
\end{table}

\section{Stellar abundances}

Abundance analysis has been performed by matching the synthetic profile with 
the observed spectrum. The line synthesis code is an updated version of the code by Monique 
Spite (1967) (e.g. Cayrel et al. 2001; Barbuy et al. 2003). The line list and atomic 
references are given in Table 3. Hyperfine structure (HFS) has been applied for 
La, Ba, Eu and Pb. The references for the HFS are: Rutten (1978) for Ba II,
 Lawler et al. (2001a) for La II, Lawler et al. (2001b) for Eu II, and Bi\'emont et al. 
(2000) for Pb I. 

\subsection{Uncertainties}

Uncertainties on abundances were calculated for both HD 11397 and HD 14282, 
by verifying how much the variation of 1$\sigma$ of the atmospheric parameters
affects the output value of the
synthesis program, here $\log{A_p}$, and also by taking into account the 
standard deviation of the abundances for which more than 2 lines
are available. Table \ref{errab} shows the values for this calculation.

Under the simplifying hypothesis of independent errors, the
uncertainty of the output value is given by

\begin{equation}
\label{erapinst}
\sigma_{Ap}=\sqrt{(\Delta A_T)^2+(\Delta A_{mt})^2+(\Delta A_l)^2+(\Delta A_\xi)^2+(sdm)^2},
\end{equation}
where $\Delta A_T$, $\Delta A_{mt}$, $\Delta A_l$, and $\Delta A_\xi$, are the 
differences in $A_p$ due to the uncertainties in the temperature, metallicity, 
$\log g$, and microturbulent velocity respectively and '$sdm$' is the standard deviation of the average.

The average value of $A_p$ ($A_{pm}$) is obtained by averaging the
individual abundances of several lines. Applying a
propagation of errors and taking into account the uncertainty
calculated with Eq. \ref{erapinst}, the uncertainty on
$A_{pm}$ is
\begin{equation}
\sigma_{Apm}={\sigma_{Ap}\over \sqrt{n_l}},
\end{equation}
where $n_l$ is the number of lines for which $\Delta A_T$, $\Delta A_{mt}$, $\Delta A_l$, 
and $\Delta A_\xi$ were computed. The uncertainty on the logarithm of $A_{pm}$ is

\begin{equation}
\sigma_{\log(Apm)}={\sigma_{Apm}\over A_{pm}\ln{10}}.
\end{equation}

The abundance $\log\epsilon$(X) is related to the output of the synthesis program by
$\log\epsilon$(X) = $\log{A_{pm}}$ + [Fe/H]. Therefore, the uncertainty is

\begin{equation}
\sigma_{\log\epsilon(X)}=\sqrt{\sigma_{\log{(Apm)}}^2+\sigma_{\rm [Fe/H]}^2}.
\end{equation}

The relation between the abundance excess relative to iron [X/Fe] and the
output value of the synthesis program is
[X/Fe] = $\log{A_{pm}}$ - $\log\epsilon_\odot(X)$, where $\log\epsilon_\odot(X)$
is the solar abundance of the element ``X''. The uncertainty is calculated by

\begin{equation}
\sigma_{\rm [X/Fe]}=\sqrt{\sigma_{\log(Apm)}^2+\sigma_{\log\epsilon_\odot(X)}^2}.
\end{equation}

For [X1/X2] the uncertainties are determined by
\begin{equation}
\sigma_{\rm [X1/X2]}=\sqrt{\sigma_{\rm [X1/Fe]}^2+\sigma_{\rm [X2/Fe]}^2},
\end{equation}

Uncertainties on elements are shown in parenthesis in Tab. \ref{medxfe} and by the error bars
in figures. 

\begin{table}
\caption{Uncertainties on abundances for HD 11397 and HD 14282.
'n' is the number of lines used to calculate the abundances;
$\lambda$ is the line used to compute the uncertainties; and 
$\log{A_{pf}}$: output with the atmospheric parameters adopted;
$\log{A_{pT}}$: output by altering the adopted $T\rm _{eff}$ by 100K;
$\log{A_{pmt}}$: output by altering +0.15 dex on adopted metallicity;
$\log{A_{pl}}$: output by altering +0.2 dex on adopted $\log g$.
$\log{A_{p\xi}}$: output by altering +0.15 kms$^{-1}$ on adopted microturbulent velocity $\xi$;
$\log(sdm)$ logarithm of the standard deviation of the average of $A_p$, called $A_{pm}$.
Uncertainties on $\log\epsilon$(X) and [X/Fe] are shown in Tab. \ref{medxfe}.}
\label{errab}
\setlength\tabcolsep{1.5pt}
\begin{tabular}{rrlcccccr}
\hline\hline
el & n & $\lambda$ ($\rm \AA$) & $\log{A_{pf}}$ & $\log{A_{pT}}$ & $\log{A_{pmt}}$ & $\log{A_{pl}}$ &
$\log{A_{p\xi}}$ & $\log(sdm)$ \\
\noalign{\smallskip}
\hline
\multicolumn{8}{c}{HD 11397} \\
\hline
C  &  3 & 5135.600 & 8.86 & 8.91 & 9.01 & 8.86 & 8.86 & ...  \\
N  &  5 & 6478.400 & 8.22 & 8.37 & 8.52 & 8.22 & 8.22 & ...  \\
Sr &  2 & 4161.79  & 3.59 & 3.62 & 3.76 & 3.64 & 3.57 & ...  \\
Y  & 12 & 5123.21  & 2.86 & 2.91 & 3.01 & 2.91 & 2.81 & 1.37 \\
Zr &  4 & 4317.30  & 3.22 & 3.27 & 3.37 & 3.30 & 3.21 & 2.53 \\
Mo &  1 & 5570.439 & 2.24 & 2.34 & 2.39 & 2.24 & 2.24 & ...  \\
Ru &  2 & 4080.594 & 2.54 & 2.69 & 2.73 & 2.54 & 2.54 & ...  \\
Ba &  5 & 5853.69  & 3.18 & 3.23 & 3.33 & 3.18 & 3.08 & 1.95 \\
La &  8 & 4123.22  & 1.88 & 1.93 & 2.03 & 1.93 & 1.83 & 0.96 \\
Ce & 10 & 4562.36  & 2.45 & 2.50 & 2.60 & 2.50 & 2.40 & 1.20 \\
Pr &  3 & 5220.11  & 1.41 & 1.46 & 1.63 & 1.51 & 1.40 & 0.93 \\
Nd &  9 & 4462.98  & 2.30 & 2.35 & 2.45 & 2.36 & 2.25 & 1.05 \\
Sm &  5 & 4318.93  & 1.36 & 1.41 & 1.51 & 1.41 & 1.35 & 0.66 \\
Eu &  2 & 6645.12  & 1.07 & 1.09 & 1.23 & 1.16 & 1.07 & ...  \\
Gd &  3 & 4085.57  & 1.42 & 1.47 & 1.62 & 1.47 & 1.42 & 0.51 \\
Dy &  2 & 4073.12  & 1.60 & 1.65 & 1.75 & 1.70 & 1.60 & ...  \\
Hf &  2 & 4080.437 & 1.83 & 1.88 & 2.01 & 1.88 & 1.81 & ...  \\
Pb &  1 & 4057.81  & 2.55 & 2.75 & 2.80 & 2.50 & 2.50 & ...  \\
\noalign{\smallskip}
\hline\hline
\multicolumn{8}{c}{HD 14282} \\

\hline
el & n & $\lambda$ ($\rm \AA$) & $\log{A_{pf}}$ & $\log{A_{pT}}$ & $\log{A_{pmt}}$ & $\log{A_{pl}}$ &
$\log{A_{p\xi}}$ & $\log(sdm)$ \\
\noalign{\smallskip}
\hline
C  &  4  & 5135.600 & 8.85 & 8.94 & 9.00 & 8.83 & 8.85 &  7.66 \\
N  &  3  & 6478.400 & 8.02 & 8.22 & 8.32 & 7.97 & 8.02 &  6.96 \\
Sr &  2  & 4161.79  & 3.67 & 3.72 & 3.82 & 3.72 & 3.62 &  ...  \\
Y  & 11  & 5123.21  & 2.89 & 2.94 & 3.04 & 2.94 & 2.84 &  1.81 \\
Zr &  4  & 4317.30  & 3.10 & 3.15 & 3.25 & 3.16 & 3.09 &  1.91 \\
Mo &  1  & 5570.439 & 2.32 & 2.42 & 2.47 & 2.37 & 2.32 &  ...  \\
Ru &  2  & 4080.594 & 2.14 & 2.20 & 2.29 & 2.19 & 2.14 &  ...  \\
Ba &  5  & 5853.69  & 2.63 & 2.70 & 2.78 & 2.68 & 2.53 &  1.65 \\
La &  4  & 4123.22  & 1.33 & 1.38 & 1.48 & 1.40 & 1.31 & -0.19 \\
Ce & 10  & 4562.36  & 1.90 & 1.95 & 2.05 & 1.97 & 1.87 &  0.74 \\
Pr &  2  & 5220.11  & 1.06 & 1.14 & 1.26 & 1.13 & 1.06 &  ...  \\
Nd &  8  & 4462.98  & 1.40 & 1.45 & 1.55 & 1.46 & 1.40 & -0.09 \\
Sm &  5  & 4318.93  & 1.01 & 1.06 & 1.08 & 1.09 & 1.01 &  0.24 \\
Eu &  2  & 6645.12  & 0.82 & 0.85 & 0.97 & 0.90 & 0.82 &  ...  \\
Gd &  3  & 4085.57  & 1.22 & 1.25 & 1.37 & 1.27 & 1.22 &  0.16 \\
Dy &  2  & 4073.12  & 1.10 & 1.15 & 1.25 & 1.17 & 1.10 &  ...  \\
Hf &  2  & 4080.437 & 1.18 & 1.23 & 1.33 & 1.25 & 1.18 &  ...  \\
Pb &  1  & 4057.81  & 2.65 & 2.80 & 2.90 & 2.65 & 2.65 &  ...  \\
\noalign{\smallskip}
\hline
\end{tabular}
\\
\end{table}

\subsection{Analysis and results}
Abundances for the sample stars are shown in Figs. 1 to 4. In order to compare our stars with stars of different classifications, we have also added strong and mild Ba-stars to the plots, and normal field stars from Boyarchuk et al. (2002), \citet{lia03}, \citet{yush04}, Allen \& Barbuy (2006) and Allen \& Porto de Mello (in preparation).

{\it Carbon and Nitrogen:}
Carbon and N abundances are plotted in Fig. 1 as a function of [Fe/H] and log g. 
As can be seen in this figure, C is slightly enhanced in both HD 11397 and HD 14282 
relative to normal stars, matching the Ba stars distributions in both plots, while 
Nitrogen seems normal for the two stars. Compared with the [C/H] vs. [Fe/H] 
relation of unpolluted stars from Fig. 11 of Masseron et al. (2006) we also found 
that both are slightly overabundant, although they can be seemingly considered as 
normal stars in this plot.

\begin{figure*}
\centering
\includegraphics[width=12cm]{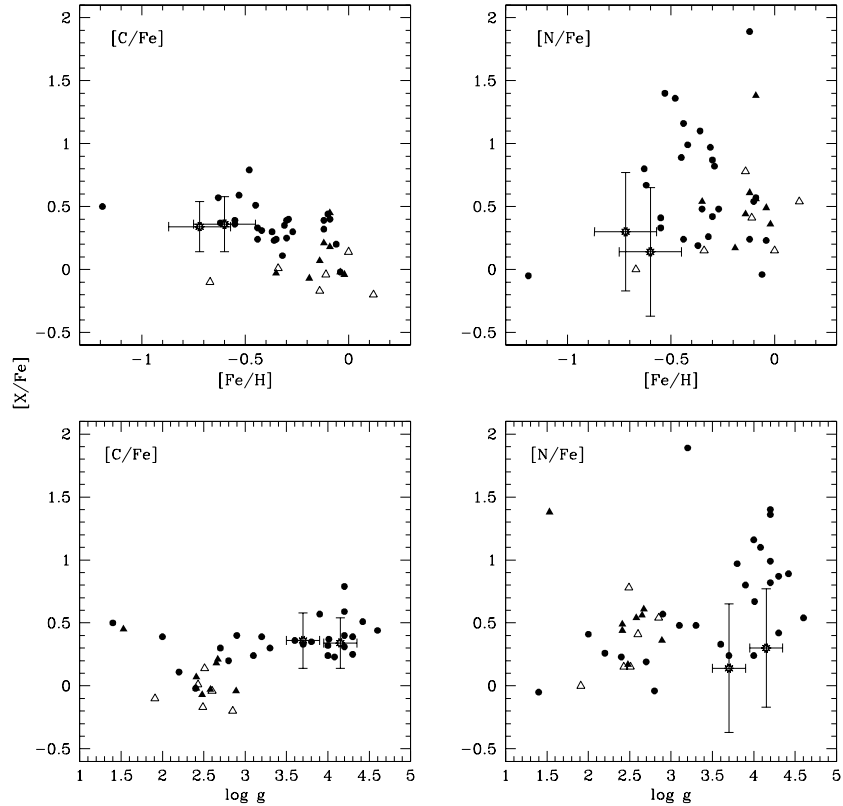}
\caption{upper panels: [C/Fe] vs. [Fe/H] and [N/Fe] vs. [Fe/H] for the sample stars.
lower panels: [C/Fe] and [N/Fe] vs. log g. Symbols: starred circles with error bars: this work; filled circles: Allen \& Barbuy 2006; filled triangles: barium stars of Allen \& Porto de Mello 2008; open triangles: stars considered normal rather than barium stars by Allen \& Porto 
de Mello (2008).
}
\label{cnfeg}
\end{figure*}

 \begin{figure*}
 \centering
 \includegraphics[width=12cm]{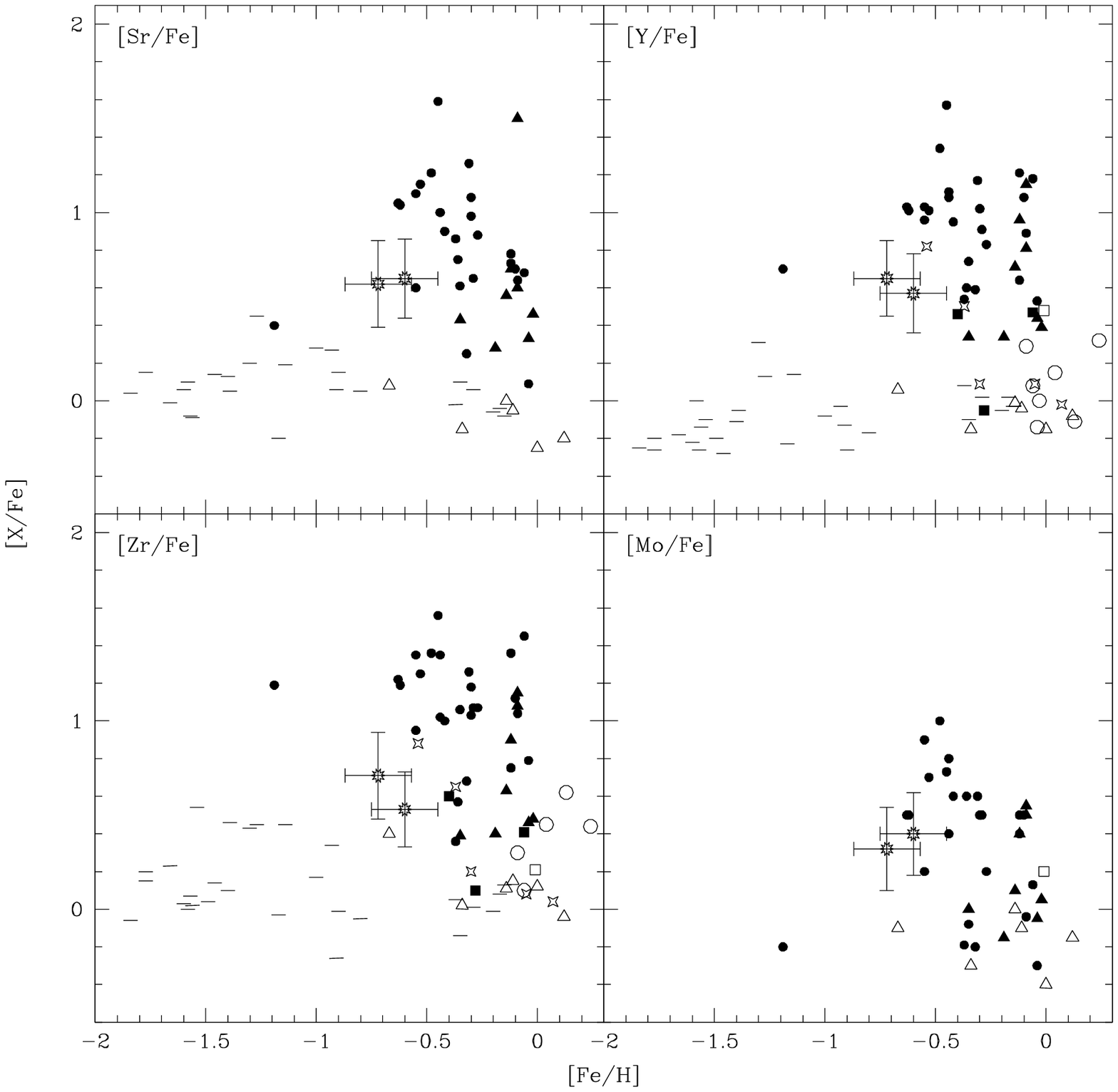}
 \caption{[X/Fe] vs. [Fe/H] for the sample stars. 
 Symbols: starred circles with error bars: this work; filled circles:  Allen \& Barbuy 2006; 
 filled triangles: barium stars of Allen \& Porto de Mello 2008; open triangles: stars considered
 normal rather than barium stars by Allen \& Porto de Mello 2008; open circles: stars from \citet{boy02};  filled squares: stars from \citet{lia03}; open squares: stars from \citet{yush04}; starred squares: barium stars from \citet{edv93}; dashes: normal stars from \citet{burr00}, \citet{edv93}, \citet{g94}, \citet{jehin99}, \citet{mg01}, \citet{tl99}, and \citet{woolf95}.}
 \label{leves}
 \end{figure*}

{\it Light $s$-process elements - Y, Zr, Sr and Mo:}
Among the light $s$-elements (hereafter $ls$) or first $s$-peak elements, 
Y is the purest, with a $s$-contribution of 92\% according to Arlandini et al. (1999), 
followed by Sr and Zr, with 85\% and 83\%, respectively, and 50\% for Mo.
In Fig. 2 we can see that, except for Zr, the $ls$ elements in HD 11397 and HD 14282 are 
mildly high when compared to normal stars, with Y showing the highest value in both stars 
($\sim$ +0.6 dex). HD 14282 shows only slight enhancements for the other $ls$-elements. 

{\it Heavy $s$-process elements - Ba, La, Ce, Nd and Hf:}
In Figs. 3 and 4, the distributions for the heavy $s$-process elements (herafter $hs$) 
are depicted. For these elements we have found a clear distinction between the two
stars, with HD 11397 showing high Ba, Ce, La, Hf, and Nd abundances compared to normal stars with  
similar metallicities. HD 14282 depicts a normal behavior with only a slight enhancement in its [Ba/Fe] ratio. Ba is the $hs$-element or second $s$-peak element (see Busso et al. 1999, 2001) with the highest $s$-contribution (81\%), followed by Ce (77\%) and La (62\%) according to \citet{arl99}. Such elements are clearly above the normal stars distribution level in HD 11397.

{\it The termination of the $s$-process path - Pb:}
Lead is at the termination of the $s$-process path, or the third peak of the $s$-process. The $s$-contribution for Pb is still uncertain. Theoretical estimations predict 46\% of yields from
the $s$-process (Arlandini et al. 1999), but observations for very low-metallicity stars indicate a strong $s$-contribution, although some contribution from the $r$-process may still be present (e.g.  Van Eck et al. 2003, Aoki et al. 2000). Lead seems normal in both HD 11397 and HD 14282.

\begin{figure*}
\centering
\includegraphics[width=12cm]{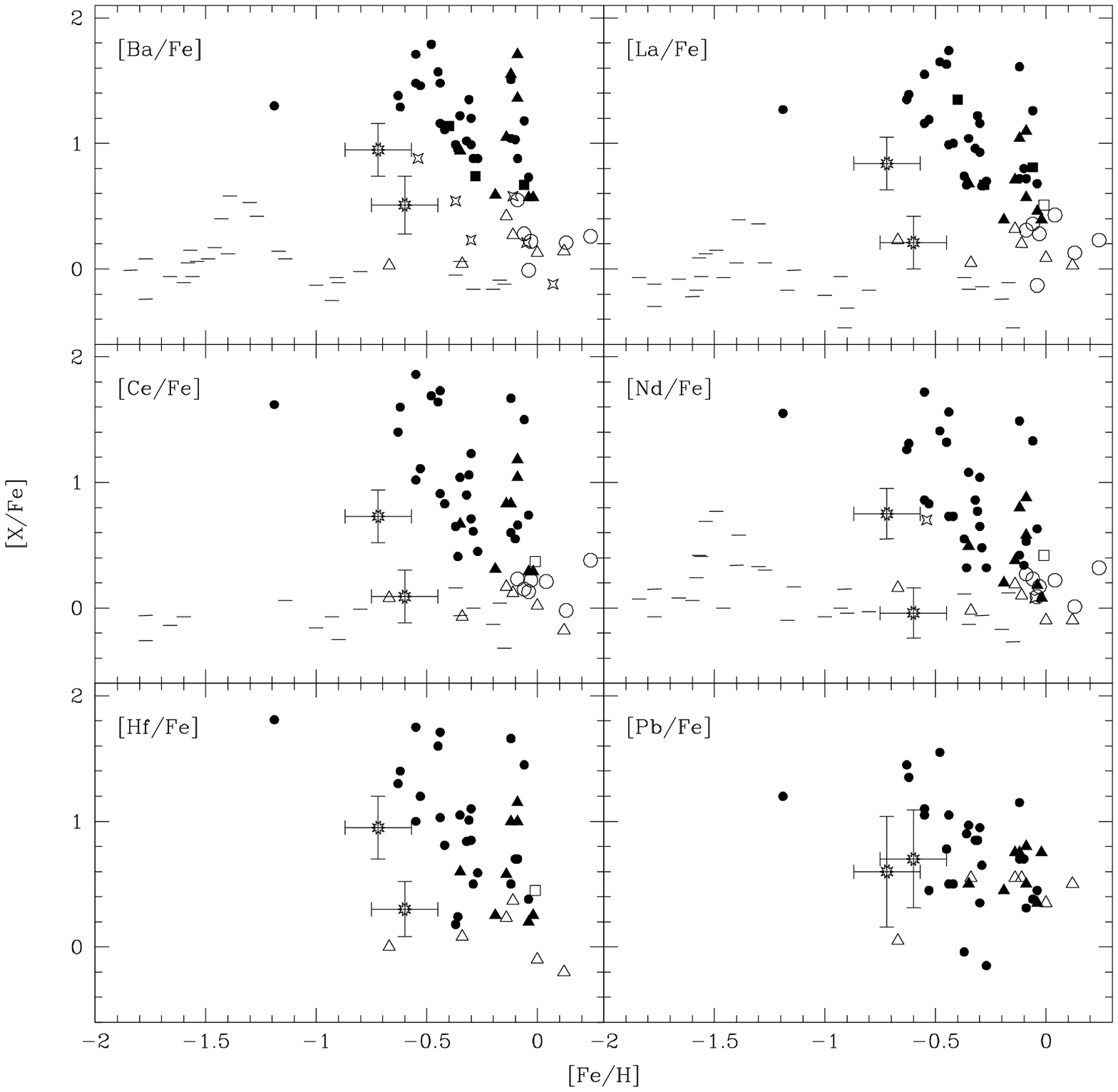}
\caption{[X/Fe] vs. [Fe/H] for the sample stars. Symbols are the same as in Fig. \ref{leves}}
\label{pesad}
\end{figure*}

\begin{figure*}
\centering
\includegraphics[width=12cm]{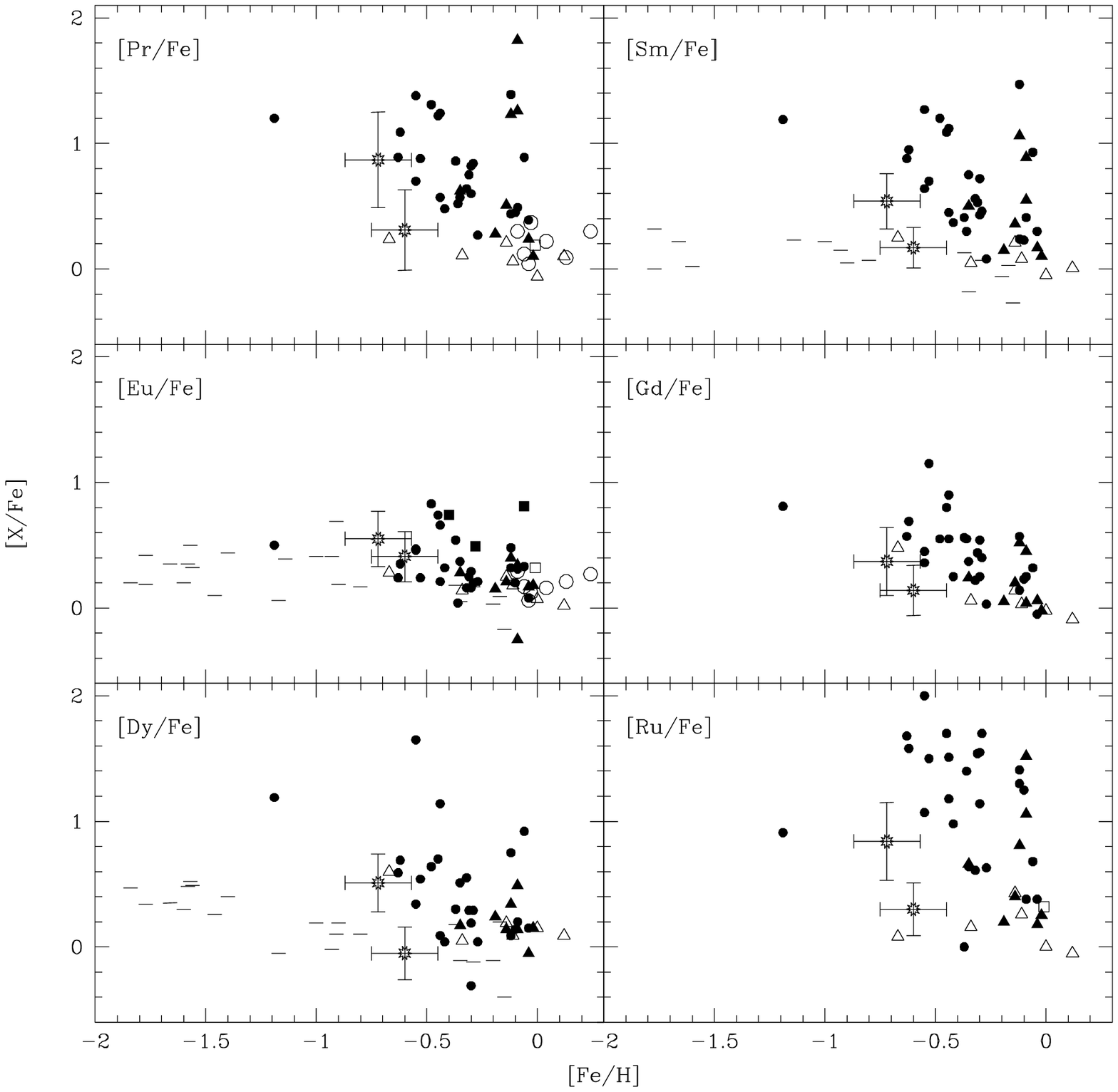}
\caption{[X/Fe] vs. [Fe/H] for the sample stars. Symbols are the same as in Fig. \ref{leves}}
\label{pesad2}
\end{figure*}

{\it The mild $s$ and $r$-elements - Pr, Ru and Sm} 
Praseodymium, Sm and Ru have a stronger $r$-contribution, with 51\% for Pr, 67\% for Sm, 
while Ru has 59\%, but they also have an $s$-contribution of $\sim$ 30\% for Ru and Sm and 49\% 
for Pr. Ru is nearer the 
first $s$-peak while Sm and Pr are nearer the second $s$-peak. These elements also
show slight enhancements in HD 11397, while HD 14282 seems to have values similar
to those of a normal star.

{\it The $r$-process dominated elements - Dy, Gd and Eu:}

Dysprosium, Gd and Eu are dominated by the $r$-process production. In Fig. 4 
their distributions are depicted and we see that HD 11397 and HD 14282 show a marginally 
underabundant behavior for Gd and Dy, while Eu abundances overlap the lower envelope of 
the Ba-stars and the normal stars distributions, indicating a small $r$-contribution for 
these stars.


\section{Discussion}
In order to improve our analysis on the abundance profile of the $s$-process elements in
HD 11397 and HD 14282, we have compared our data with theoretical surface abundances of AGB stars from Goriely \& Mowlavi (2000) in Fig. \ref{mod_comp}. If those stars inherited their abundance
content from an AGB star, their surface $s$-elements abundances should mimic those of an AGB star.
The models are for stars with metallicities similar to our sample stars. Solid lines represent surface abundances after 10 dredge-ups, and dotted lines, after 30 dredge-ups. The upper panel shows that most neutron-capture elements in HD 14282 are below the abundances predicted for a 10 dredge-up model. Therefore that star seems to show a slight abundance anomaly for some neutron-capture elements (Y, Sr, Mo, Pb) while other elements are normal. Boyarchuk et al. (2002) have also found some slight $s$-elements anomaly in normal field red giants. If these chemical anomalies in HD 14282 were inherited from the proto-cloud of the star, such excess should be ascribed to a pristine contamination. If they are due to the beginning of the $s$-process operation and the dredge-up of material enriched in carbon and $s$-elements, HD 14282 $s$-anomaly may be ascribed to a mass-transfer mechanism.

The abundances of HD 11397 for most of the elements in Fig. \ref{mod_comp} are between the 
10 and the 30 dredge-ups curves, in agreement with an AGB abundance profile.
Compared to normal disk stars, HD 11397 shows an overabundant behavior for most of the neutron capture elements and seems to share its chemical profile with the mild Ba stars. We suggest that this is also a mild Ba-star. An apparent higher $hs$ content than the $ls$-elements is seen in that star. For both HD 11397 and HD 14282, the heavy element Pb, is underabundant relative to these models predictions. 

\begin{figure*}
\includegraphics[width=6cm]{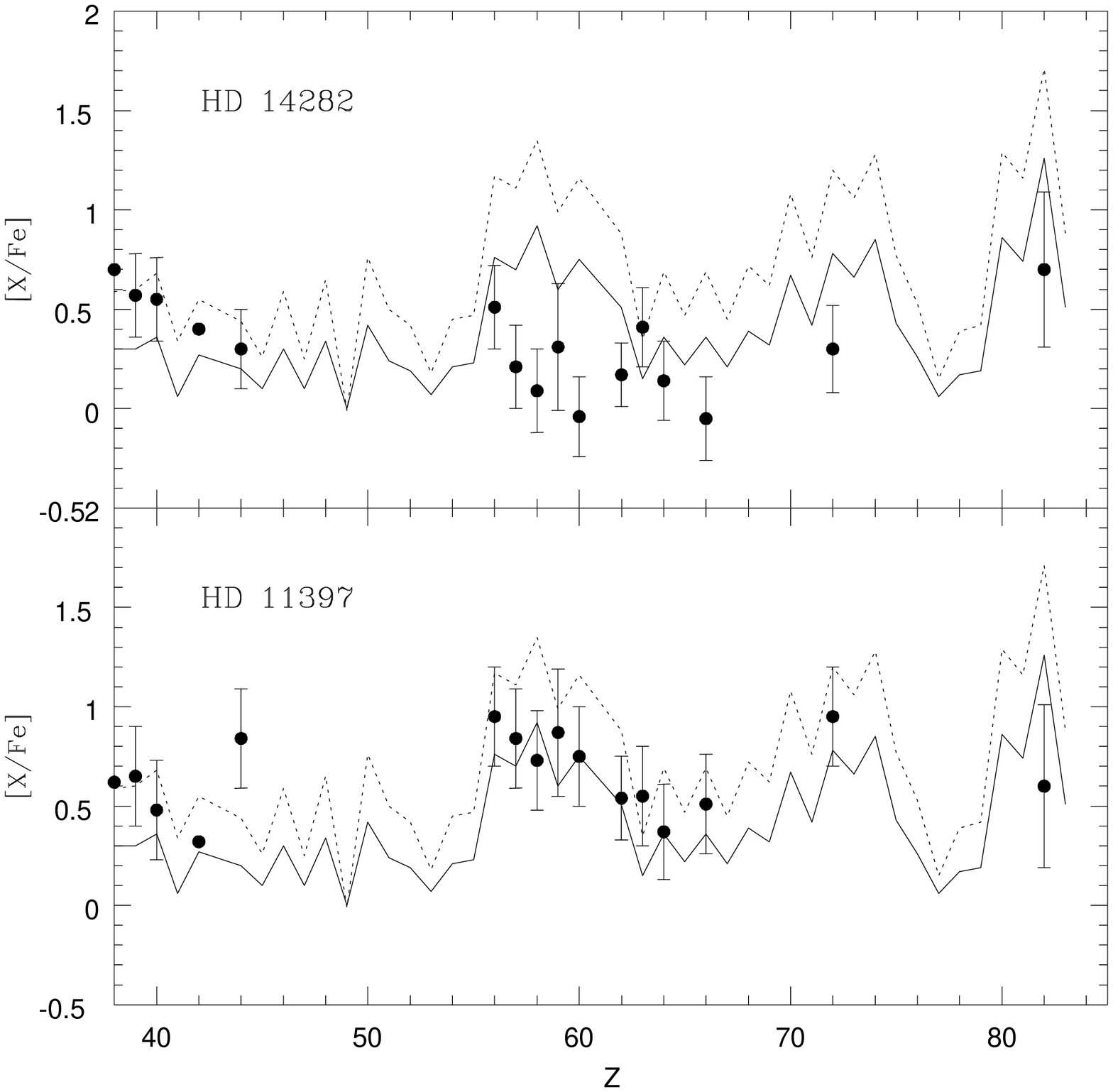}
\caption{[X/Fe] vs. Z for HD 14282 (upper panel) and HD 11397 (lower panel), 
compared to theoretical surface abundances predicted for AGB stars 
of Goriely \& Mowlavi 2000. Our data are compared to models with similar 
metallicities to the sample stars. Solid lines represent surface abundances 
predicted for 10 dredge-ups and dotted lines, for 30 dredge-ups.
}
\label{mod_comp}
\end{figure*}
 Radial velocities of the bulgelike stars have been derived using CORAVEL, and possible binaries have been discarded (Grenon, 1998). Nevertheless, a detailed study of the radial velocities of HD 11397 and HD 14282 should be performed in order to have a robust statement about their single nature. If their non-binarity is confirmed, they could be used as templates for the study of $s$-enriched stars with a non-binary origin, and therefore in a scenario different from that of the mass-transfer paradigm.

\section{Summary}
We have performed a chemical abundance analysis of two dwarf stars with $s$-process 
anomalies, HD 11397 and HD 14282. We aim to define if they can be considered as Ba-stars, and, if confirmed, to infer their Ba degree, i.e., to define if they are mild or strong Ba stars.
Abundances of 18 neutron-capture elements, with different $s$ and/or $r$ processes contributions,
have been derived. The resulting abundance ratios of the two stars have been compared to those
of normal stars, to abundance ratios of mild and strong Ba stars and to theoretical predictions
for AGB stars. 
We have found that HD 11397 shows a mild enhancement for most of the $s$-process elements
as well as for some $r$-process dominated elements. This star seems to share its abundance
profile with the mild Ba-stars. Although showing some slight chemical anomalies for Y, 
Sr and Mo, HD 14282 depicts a chemical pattern similar to the normal stars with
slight $s$-process enhancements.

\begin{acknowledgements}
      Part of this work was supported by the FAPESP fellowship \#01/14594-2.
      We acknowledge FAPESP project \#98/10138-8. We warmufully thank A. Jorissen for
      his careful revision of the paper. DMA acknowledges CAPES for the post-doctoral fellowship n$^{\circ}$ BEX 3448/06-1. We are grateful to Beatriz Barbuy for making available the spectrum synthesis code.

\end{acknowledgements}


\Online

{\scriptsize 
\setlength\tabcolsep{3pt}
\begin{longtable}{ccrrlcrrrcrrr}
\caption{Abundance results for the sample stars. The G band ranges from 4295 to 4315 $\rm \AA$.}
\label{eqwid} \\

\hline\hline
 &&&& &&  \multicolumn{3}{c}{HD 11397} && \multicolumn{3}{c}{HD 14282} \\ 
\cline{7-9} \cline{11-13} \\
el & $\lambda$ &&& molec &&& $\log\epsilon$(X) & [X/Fe] &&& $\log\epsilon$(X) & [X/Fe] \\
\hline 
\endfirsthead

\multicolumn{13}{c}%
{{\bfseries \tablename\ \thetable{} -- continued}} \\
\hline\hline
 &&&& &&  \multicolumn{3}{c}{HD 11397} && \multicolumn{3}{c}{HD 14282} \\ 
\cline{7-9} \cline{11-13} \\
el & $\lambda$ &&& molec &&& $\log\epsilon$(X) & [X/Fe] &&& $\log\epsilon$(X) & [X/Fe] \\
\hline 
\endfirsthead
\hline
\endlastfoot

 C    & G band   &        &	    &	CH	   && ... &  8.14 &  0.34 && ... &  8.25 &  0.33 \\
 C    & 5135.600 &        &	    &	C2	   && ... &  8.14 &  0.34 && ... &  8.25 &  0.33 \\
 C    & 5165.254 &        &	    &	C2 Swan(0,0) && ... &  8.14 &  0.34 && ... &  8.25 &  0.33 \\
 C    & 5635.500 &        &	    &	C2 Swan(0,1) && ... & ...   & ...   && ... &  8.35 &  0.43 \\
 N    & 6477.200 &        &	    &	CN	   && ... &  7.50 &  0.30 && ... &  7.42 &  0.10 \\
 N    & 6478.400 &        &	    &	CN	   && ... &  7.50 &  0.30 && ... &  7.42 &  0.10 \\
 N    & 6478.700 &        &	    &	CN	   && ... &  7.50 &  0.30 && ... &  7.52 &  0.20 \\
 N    & 6703.968 &        &	    &	CN	   && ... &  7.50 &  0.30 && ... & ...   & ...   \\
 N    & 6706.733 &        &	    &	CN	   && ... &  7.50 &  0.30 && ... & ...   & ...   \\
 N    & 6708.993 &        &	    &	CN	   && ... & ...   & ...   && ... & ...   & ...   \\
\noalign{\smallskip}
\hline\hline 
el & $\lambda$ & $\chi_{ex}$ & $\log$ $gf$ & ref && EW & $\log\epsilon$(X) & [X/Fe] && EW & $\log\epsilon$(X) & [X/Fe] \\ 
\noalign{\smallskip}
\hline
Sr I  & 4607.340 &  0.000 &   0.280 &	G94	   &&  62 &  2.87 &  0.62 &&  57 &  3.07 &  0.70 \\
Sr I  & 6791.050 &  1.760 &  -0.850 &	G94	   && ... & ...   & ...   && ... & ...   & ...   \\
Sr II & 4077.710 &  0.000 &   0.170 &	G94        && 274 &  2.87 &  0.62 && 165 &  2.97 &  0.60 \\
Sr II & 4161.790 &  2.940 &  -0.600 &	G94	   &&  46 &  2.87 &  0.62 &&  49 &  3.07 &  0.70 \\
 Y II & 4883.690 &  1.080 &   0.070 &	H82	   &&  75 &  2.24 &  0.72 &&  82 &  2.49 &  0.85 \\
 Y II & 4982.140 &  1.030 &  -1.290 &	H82	   &&  16 &  2.14 &  0.62 &&  18 &  2.09 &  0.45 \\
 Y II & 5087.430 &  1.080 &  -0.170 &	H82	   &&  59 &  2.14 &  0.62 &&  67 &  2.19 &  0.55 \\
 Y II & 5119.110 &  0.990 &  -1.360 &	H82	   &&  34 &  2.19 &  0.67 &&  26 &  2.09 &  0.45 \\
 Y II & 5123.210 &  0.990 &  -0.830 &	H82	   &&  51 &  2.14 &  0.62 &&  52 &  2.29 &  0.65 \\
 Y II & 5200.420 &  0.990 &  -0.570 &	H82	   &&  56 &  2.19 &  0.67 &&  57 &  2.19 &  0.55 \\
 Y II & 5205.720 &  1.030 &  -0.340 &	H82	   &&  59 &  2.09 &  0.57 &&  60 &  2.19 &  0.55 \\
 Y II & 5289.820 &  1.030 &  -1.850 &	H82	   &&  11 &  2.14 &  0.62 &&   8 &  2.09 &  0.45 \\
 Y II & 5402.780 &  1.840 &  -0.520 &	HL83	   &&  23 &  2.14 &  0.62 &&  25 &  2.09 &  0.45 \\
 Y II & 5473.390 &  1.740 &  -1.020 &	H82	   &&  12 &  2.14 &  0.62 &&  12 &  2.19 &  0.55 \\
 Y II & 5728.890 &  1.840 &  -1.120 &	H82	   &&	7 &  2.24 &  0.72 && ... & ...   & ...   \\
 Y II & 6795.410 &  1.730 &  -1.250 &	M94	   &&  10 &  2.19 &  0.67 &&  22 &  2.19 &  0.55 \\
Zr I  & 6127.470 &  0.150 &  -1.060 &	B81	   &&	9 &  2.50 &  0.62 &&   5 &  2.60 &  0.60 \\
Zr I  & 6134.570 &  0.000 &  -1.280 &	B81	   &&	7 &  2.30 &  0.42 && ... & ...   & ...   \\
Zr I  & 6140.460 &  0.520 &  -1.410 &	B81	   &&	2 &  2.30 &  0.42 && ... & ...   & ...   \\
Zr I  & 6143.180 &  0.070 &  -1.100 &	B81	   &&	6 &  2.30 &  0.42 && ... & ...   & ...   \\
Zr I  & 6489.650 &  1.550 &   0.250 &	T90,K85    && ... & ...   & ...   &&   3 &  2.50 &  0.50 \\
Zr II & 4208.980 &  0.710 &  -0.460 &	B81	   &&  59 &  2.40 &  0.52 &&  60 &  2.50 &  0.50 \\
Zr II & 4317.300 &  0.710 &  -1.380 &	B81	   &&  32 &  2.50 &  0.62 &&  28 &  2.50 &  0.50 \\
Zr II & 5350.090 &  1.830 &  -0.930 &	T89,K85    &&  10 &  2.70 &  0.82 &&  11 &  2.50 &  0.50 \\
Zr II & 6114.800 &  1.670 &  -1.700 &	S00	   &&	6 &  2.70 &  0.82 &&   9 &  2.60 &  0.60 \\
Mo I  & 5570.439 &  1.330 &   0.150 &	S00	   &&  13 &  1.52 &  0.32 &&   4 &  1.72 &  0.40 \\
Ru I  & 4080.594 &  0.810 &  -0.040 &	VALD	   &&	9 &  1.82 &  0.70 &&   2 &  1.54 &  0.30 \\
Ru I  & 4757.856 &  0.928 &  -0.540 &	ajusol     &&	8 &  2.07 &  0.95 &&   6 &  1.54 &  0.30 \\
Ba II & 4554.030 &  0.000 &   0.170 &	M98	   && 384 &  2.36 &  0.95 && 198 &  1.93 &  0.40 \\
Ba II & 4934.100 &  0.000 &  -0.150 &	M98	   && 192 &  2.26 &  0.85 && 179 &  1.93 &  0.40 \\
Ba II & 5853.690 &  0.604 &  -1.010 &	M98	   && 119 &  2.46 &  1.05 &&  72 &  2.03 &  0.50 \\
Ba II & 6141.727 &  0.704 &  -0.070 &	M98	   && 203 &  2.36 &  0.95 && 120 &  2.13 &  0.60 \\
Ba II & 6496.909 &  0.604 &  -0.380 &	R78	   && 163 &  2.36 &  0.95 && 107 &  2.13 &  0.60 \\
La II & 4086.710 &  0.000 &  -0.070 &	LS01	   &&  65 &  1.16 &  0.75 &&  43 &  0.73 &  0.20 \\
La II & 4123.220 &  0.320 &   0.130 &	LS01	   &&  69 &  1.16 &  0.75 &&  54 &  0.73 &  0.20 \\
La II & 5123.000 &  0.320 &  -0.850 &	LS01	   &&  20 &  1.21 &  0.80 &&   7 &  0.73 &  0.20 \\
La II & 5303.530 &  0.320 &  -1.350 &	LS01	   &&  19 &  1.41 &  1.00 &&   7 &  0.78 &  0.25 \\
La II & 5797.570 &  0.240 &  -1.360 &	LS01	   &&  10 &  1.26 &  0.85 && ... & ...   & ...   \\
La II & 5805.770 &  0.120 &  -1.560 &	LS01	   &&  13 &  1.16 &  0.75 && ... & ...   & ...   \\
La II & 5863.710 &  0.930 &  -1.370 &	LS01	   &&	4 &  1.41 &  1.00 && ... & ...   & ...   \\
La II & 6390.492 &  0.320 &  -1.410 &	LS01	   &&  11 &  1.16 &  0.75 && ... & ...   & ...   \\
Ce II & 4073.470 &  0.480 &   0.180 &	PQ00	   &&  31 &  1.63 &  0.65 &&  16 &  1.10 &  0.00 \\
Ce II & 4120.830 &  0.320 &  -0.290 &	PQ00	   &&	2 &  1.63 &  0.65 &&  13 &  1.10 &  0.00 \\
Ce II & 4145.000 &  0.700 &   0.090 &	PQ00	   &&  17 &  1.63 &  0.65 &&   8 &  1.10 &  0.00 \\
Ce II & 4222.600 &  0.120 &  -0.090 &	PQ00       &&  38 &  1.63 &  0.65 &&  16 &  0.90 & -0.20 \\
Ce II & 4418.780 &  0.860 &   0.230 &	PQ00	   &&  40 &  1.73 &  0.75 &&  20 &  1.30 &  0.20 \\
Ce II & 4486.910 &  0.300 &  -0.330 &	PQ00	   &&  38 &  1.83 &  0.85 &&  17 &  1.30 &  0.20 \\
Ce II & 4523.070 &  0.520 &  -0.080 &	PQ00	   &&  34 &  1.63 &  0.65 &&  20 &  1.30 &  0.20 \\
Ce II & 4562.360 &  0.480 &   0.190 &	PQ00	   &&  42 &  1.73 &  0.75 &&  24 &  1.30 &  0.20 \\
Ce II & 4628.160 &  0.520 &   0.150 &	PQ00	   &&  42 &  1.73 &  0.75 &&  21 &  1.10 &  0.00 \\
Ce II & 5330.560 &  0.870 &  -0.510 &	PQ00,K85   &&  12 &  1.78 &  0.80 &&   4 &  1.20 &  0.10 \\
Ce II & 5975.820 &  1.330 &  -0.530 &	PQ00	   &&  10 &  1.83 &  0.85 && ... & ...   & ...   \\
Pr II & 5220.110 &  0.800 &   0.160 &	G91,L76    &&  13 &  0.69 &  0.75 &&  17 &  0.46 &  0.40 \\
Pr II & 5259.730 &  0.630 &   0.080 &	G91,L76    &&  11 &  0.69 &  0.75 &&   2 &  0.26 &  0.20 \\
Pr II & 5352.410 &  0.480 &  -0.810 &	T89,K85    &&	5 &  0.99 &  1.05 && ... & ...   & ...   \\
Nd II & 4018.820 &  0.060 &  -0.850 &	H03	   &&  24 &  1.38 &  0.65 &&  15 &  0.80 & -0.05 \\
Nd II & 4021.330 &  0.320 &  -0.100 &	H03	   &&  36 &  1.43 &  0.70 &&  12 &  0.80 & -0.05 \\
Nd II & 4061.080 &  0.470 &   0.550 &	MW77	   &&  63 &  1.63 &  0.90 &&  54 &  0.80 & -0.05 \\
Nd II & 4446.380 &  0.200 &  -0.350 &	S96	   &&  30 &  1.38 &  0.65 &&  14 &  0.80 & -0.05 \\
Nd II & 4462.980 &  0.560 &   0.040 &	MW77	   &&  40 &  1.58 &  0.85 &&  19 &  0.80 & -0.05 \\
Nd II & 5130.590 &  1.300 &   0.450 &	H03	   &&  30 &  1.43 &  0.70 &&  12 &  0.80 & -0.05 \\
Nd II & 5319.820 &  0.550 &  -0.140 &	H03	   &&  29 &  1.43 &  0.70 &&  10 &  0.80 & -0.05 \\
Nd II & 5688.530 &  0.990 &  -0.310 &	H03	   &&  14 &  1.48 &  0.75 &&   8 &  0.90 &  0.05 \\
Nd II & 5740.870 &  1.160 &  -0.530 &	H03	   &&	7 &  1.48 &  0.75 && ... & ...   & ...   \\
Sm II & 4318.930 &  0.280 &  -0.270 &	B89	   &&  29 &  0.64 &  0.35 &&  30 &  0.41 &  0.00 \\
Sm II & 4499.470 &  0.250 &  -1.010 &	B89	   &&	9 &  0.94 &  0.65 &&   5 &  0.61 &  0.20 \\
Sm II & 4537.940 &  0.480 &  -0.230 &	B89	   &&  21 &  0.79 &  0.50 &&  13 &  0.61 &  0.20 \\
Sm II & 4577.690 &  0.250 &  -0.770 &	B89	   &&  10 &  0.94 &  0.65 &&   7 &  0.71 &  0.30 \\
Sm II & 4815.800 &  0.180 &  -0.770 &	B89	   &&  11 &  0.74 &  0.45 &&   7 &  0.51 &  0.10 \\
Eu II & 6437.700 &  1.320 &  -0.320 &	LD01	   &&  10 &  0.35 &  0.55 &&  10 &  0.42 &  0.50 \\
Eu II & 6645.120 &  1.380 &   0.120 &	LD01	   &&	7 &  0.35 &  0.55 &&   9 &  0.22 &  0.30 \\

\hline\hline 
el & $\lambda$ & $\chi_{ex}$ & $\log$ $gf$ & ref && EW & $\log\epsilon$(X) & [X/Fe] && EW & $\log\epsilon$(X) & [X/Fe] \\ 
\noalign{\smallskip}
\hline

Gd II & 4037.910 &  0.560 &  -0.230 &	B88	   &&  10 &  0.85 &  0.45 &&   4 &  0.62 &  0.10 \\
Gd II & 4085.570 &  0.730 &   0.070 &	B88	   &&	6 &  0.70 &  0.30 &&   5 &  0.62 &  0.10 \\
Gd II & 4191.080 &  0.430 &  -0.680 &	C62	   &&  10 &  0.75 &  0.35 &&   4 &  0.72 &  0.20 \\
Dy II & 4073.120 &  0.540 &  -0.330 &	K92-BL93m  &&	6 &  0.88 &  0.40 &&   5 &  0.50 & -0.10 \\
Dy II & 4103.310 &  0.100 &  -0.370 &	K92-BL93m  &&  29 &  1.08 &  0.60 &&  63 &  0.60 &  0.00 \\
Hf II & 4080.437 &  0.608 &  -1.050 &	ajusol     &&  15 &  1.11 &  0.95 &&  10 &  0.58 &  0.30 \\
Hf II & 4093.155 &  0.452 &  -1.090 &	VALD	   &&  21 &  1.11 &  0.95 &&   5 &  0.58 &  0.30 \\
Pb I  & 4057.810 &  1.320 &  -0.220 &	B00        && 111 &  1.83 &  0.60 &&  34 &  2.05 &  0.70 \\
\noalign{\smallskip}
\hline
\end{longtable}

$\log\epsilon$(X)=($\log{n_X/n_H}$)+12 and [X/Fe]=$\log\epsilon$(X)$_\ast$-$\log\epsilon$(X)$_\odot$-[Fe/H]
}

\end{document}